\documentclass[12pt,a4paper]{article}

\usepackage[utf8]{inputenc}
\usepackage{amsmath}
\usepackage{amssymb}
\usepackage{bm}

\usepackage{bm}
\usepackage{url}
\usepackage{tikz}
\makeatletter
\renewcommand{\fnum@figure}{Fig. \thefigure}
\makeatother

\title{\bf On scattering problem off the potential, decreasing as inverse square of distance
}

\author{{V.\,A. Gradusov\/\thanks{e-mail: v.gradusov@spbu.ru}\footnote{St Petersburg state university, St Petersburg 199034, Russia}, S.\,L. Yakovlev\/\thanks{e-mail: s.yakovlev@spbu.ru}  }\footnote{St Petersburg state university, St Petersburg 199034, Russia}}

\begin{document}
\date{}	
\maketitle

\abstract{A solution of the scattering problem  is obtained  for the Schrödinger equation with the potential of induced dipole interaction, which decreases as the inverse square of the distance. Such a potential arises in the collision of an incident charged particle with a complex of charged particles (for example, in the collision of electrons with atoms). For the wave function, an integral equation is constructed for an arbitrary value of the orbital momentum of relative motion. By solving this equation, an exact integral representation for the $K$-matrix of the problem is obtained in terms of the wave function. This representation is used to analyze the behavior of the $K$-matrix at low energies and to obtain comprehensive information on its threshold behavior for various values of the dipole momentum. The resulting solution is applied to study the behavior of the scattering cross sections in the electron, positron and antiproton system.	
 }



\section{Introduction}

The interaction between a charged particle and a target containing charged particles leads to the appearance of a slowly decreasing potential at large distances from the target, which has the form of the sum of multipole terms \cite{Landau}. The leading terms in this expansion are the Coulomb interaction (in the case of a charged target) and the induced dipole interaction, which is inversely proportional to the square of the distance $ \pm \alpha^2 r^{-2}$. The next expansion terms, decreasing as $O(r^{-3})$ and faster, do not contribute to the behavior of the leading terms of the wave functions and can be taken into account by perturbation theory methods when solving the scattering problem.
The induced dipole interaction appears when the state of the target is not spherically symmetric, as, for example, in the scattering of electrons by excited states of the hydrogen atom. In the theoretical solution of the scattering problem, such an interaction is obtained as a result of expanding the wave function of a many-particle system in terms of the wave functions of the target states and, in this sense, is an effective potential for effective equations describing the relative motion of an incident particle and a target \cite{ZhZa, Taylor}.
The motion in the $+\alpha^2r^{-2}$ repulsive field is quite regular, while in the $-\alpha^2r^{-2}$ attractive field, motion at large distances acquires a number of features \cite{Landau}. In particular, at a sufficient  attraction intensity $\alpha^2$, characteristic oscillations may appear in  the scattering cross section just above the threshold, which were first theoretically predicted in \cite{GD-1, GD-2} for the problem of electron scattering on a hydrogen atom. The so-called computational method of $R$-matrix \cite{Desc, Burke} is traditionally used to find the behavior of the cross section near the threshold. Being useful for the numerical solution of the Schrödinger equation, this method, however, cannot be considered completely rigorous from a mathematical point of view. A rigorous approach is based on integral equations of the Lippmann-Schwinger (L-S) type, which make it possible to prove the existence and uniqueness of a solution of the scattering problem, and also  to obtain complete solutions to the scattering problem, including representations for scattering amplitudes, scattering matrices, etc. In this paper, we construct an integral equation of this type for solving the problem of scattering by a potential that decreases at large distances as
$-\alpha^2/r^2$, and study its solution at low energies. The resulting solution is then used to analyze the solution of the multichannel scattering problem for the electron, positron, antiproton system. Although it would be possible to take another system of three charged particles, however, it is this system that has recently attracted the attention of researchers, as it plays an important role in experiments aimed to study antimatter \cite{amatter-1, amatter-2}. The theoretical study of collision processes in such a three-body system at near-threshold energies is very difficult because the wave function approaches its asymptote extremely slowly
due to the slow decrease in the dipole part of the interaction \cite{Hu} - \cite{Grad3} . This, in particular, explains the relatively small number of theoretical papers devoted to the study of scattering processes in such systems of charged particles at low energies. 
Progress in the numerical solution critically depends on the use of the wave function for the problem of scattering off a potential with induced dipole interaction in the three-particle equations, which for these equations  plays the role of an asymptotic boundary condition.

It should be noted that for a long time the main theoretical interest in problems for the Schrödinger equation with a potential inversely proportional to the square of the distance was mainly associated with the three-particle Thomas and Efimov effects \cite{Tom, Efim} of the appearance of  infinite series of three-particle bound states. These effects are generated by an effective three-particle attractive  potential, inversely proportional to the square of the hyperradius, which occurs at small distances between particles for a special class of pair interactions (see, for example, one of the recent papers \cite{KM}). We should also mention the works (see \cite{Pup} and references therein), which studied three-body bound states for particles with pair interactions inversely proportional to squared distances.
The peculiarities of the dynamics of particles interacting at large distances by potentials decreasing as the inverse square of the distance have not received sufficient attention in theoretical works (the only exception is the paper \cite{Ros}). In this paper, we fill this gap and carry out a detailed study of the solution of the Schrödinger equation for the scattering problem with such potentials on the basis of integral equations of the Lippmann-Schwinger type.


\section{One channel scattering problem
}
The problem of single-channel
scattering\footnote{We have chosen the formulation of the scattering problem leading to a real solution. Other equivalent formulations are obtained by simple renormalization.}
is to find a solution to the Schrödinger equation 

\begin{equation}
\left[-\frac{d^2}{dr^2} +\frac{\ell(\ell+1)}{r^2}+V^{d}(r)-p^2 \right] \psi(r,p)=0, 
\label{SE}
\end{equation}
on an interval 	$0\le r < \infty $,  which obeys the regularity condition at $r=0$ 
\begin{equation}
\psi(0,p)=0
\label{psi=0}
\end{equation}
and  the asymptotic condition as $r\to \infty$
\begin{equation}
\psi(r,p)\sim  [\sin(pr-\ell\pi/2)+\cos(pr-\ell\pi/2)K_\ell(p)]A_\ell  
\label{psi-as}
\end{equation}
with some real amplitude $A_\ell$. Due to the homogeneity of the equation (\ref{SE}), the specific value of the constant $A_\ell$ is chosen according to the normalization of the wave function.   The real unknown quantity  $K_\ell(p)$ determined in the process of solution is called the $K$-matrix.
The real-valued potential $V^d(r)$ corresponding to the dipole interaction at large distances is given by  formulas     
\begin{equation*}
V^{d}(r)= 
\left \{ 
\begin{array}{l} 
V(r), \ \ \ 0\le r < R, \\
-\alpha^2/r^2, \ \ \ r\ge R. 
\end{array}  
\right. 
\label{V-alpha}
\end{equation*}
Here $V(r)$, which  determines the interaction at small and medium distances, is a sufficiently smooth function and satisfies the standard condition for small $r$
\begin{equation}
\int_0^c rV(r)dr  < \infty,  
\label{rV}
\end{equation}
with some  constant $c$ such that   $0<c<\infty$. 
The parameter $R$
corresponds to the distance at which $V^{d}(r)$ becomes the dipole attraction.
The potential $V^d(r)$ satisfies the conditions \cite{DeAlfaro}, which are sufficient for the existence and uniqueness of the solution of the scattering problem (\ref{SE})-(\ref{psi-as}). However, the long range nature of the 
interaction $V^d(r)=O(1/r^2)$ leads to the practical uselessness of the asymptotic condition (\ref{psi-as}), in particular, for small values of $p$. In this regard, it is required to construct a solution algorithm in which
the long-range part of the interaction is taken into account explicitly.
We implement such an algorithm using the  potential splitting method \cite{Split1}-\cite{Split4}, in which the potential
$V^{d}$ is represented as the sum of the finite-range potential $V_R$ and the long-range part $V^R$
\begin{equation*}
V^{d}(r)=V_R(r)+V^R(r).  
\label{Split-pot}
\end{equation*}
Here the $V_R$ and $V^R$ components are given by 
\begin{equation*}
V_R(r)= 
\left \{ 
\begin{array}{l} 
V(r)+V_0 
, \ \ \ 0\le r < R, \\
0, \ \ \ r\ge R,
\end{array}  
\right. 
\label{V_R}
\end{equation*}
\begin{equation*}
V^R(r)= 
\left \{ 
\begin{array}{l} 
-V_0
, \ \ \ 0\le r < R,  \\
-\alpha^2/r^2, \ \ \ r\ge R. 
\end{array}  
\right.  
\label{V_R}
\end{equation*}
In the general case, the parameter $V_0>0$ can be chosen quite arbitrarily, while the choice of $V_0=\alpha^2/R^2$ is the most natural.     
Let us now transform the equation (\ref{SE}) to the form 
\begin{equation}
\left[-\frac{d^2}{dr^2} +\frac{\ell(\ell+1)}{r^2}+V^R(r)-p^2 \right] \psi(r,p)=-V_R(r)\psi(r,p). 
\label{SE-R}
\end{equation}
We solve the equation (\ref{SE-R}) in two stages.
At first stage, let us consider the homogeneous equation with potential  $V^R$
\begin{equation}
\left[-\frac{d^2}{dr^2} +\frac{\ell(\ell+1)}{r^2}+V^R(r)-p^2 \right] \phi(r,p)=0
\label{SE-V^R}
\end{equation} 
and construct regular and non-regular solutions and respective Green's function.  
Regular solution, which satisfies the condition  $\phi(0,p)=0$, on the interval $0\le r\le R$  is given by
\begin{equation}
\phi(r,p)=a_1(p) F_\ell(qr), \ \ \ q^2=p^2+V_0, 
\label{phi_R}
\end{equation}
At  $r\ge R$ let us choose this solution in the form consistent with (\ref{psi-as}),  
\begin{equation*}
\phi(r,p)= F_\lambda(pr) + b_1(p) G_\lambda(pr). 
\label{phi^R}
\end{equation*}
Here, the non-integer in the general case $\lambda$ means the root of the equation \\ $$\ell(\ell+1)-\alpha^2=\lambda(\lambda+1),$$ that  becomes equal to $\ell$ if $\alpha=0$: 
$$
\lambda=-1/2+\sqrt{1/4+\ell(\ell+1)-\alpha^2}.
$$
The functions $F_\lambda$ and $G_\lambda$ are expressed in terms of the Bessel functions
\begin{align}
F_\lambda(z)= \sqrt{\frac{\pi z}{2}}J_{\lambda+1/2}(z), \nonumber 
\\  
G_\lambda(z)=\frac{1}{\cos \lambda \pi} \sqrt{\frac{\pi z}{2}}J_{-\lambda-1/2}(z)  \label{G}
\end{align}  
  and they  are linearly independent solutions of the equation 
\begin{equation*}
\left[-\frac{d^2}{dz^2} +\frac{\lambda(\lambda+1)}{z^2}-1 \right] F_\lambda(G_\lambda)=0
\label{SE-lambda}
\end{equation*}
at arbitrary $\lambda$, including complex value case. 
The choice of the normalization of the function $F_\lambda(z)$ is standard, however  there are various options in the literature for the irregular solution $G_\lambda(z)$. Our choice in (\ref{G}) is made in such a way that, for an integer value of $\lambda=\ell\ge 0$, the function $G_\ell(z)$ coincides with the Coulomb function
$G_\ell(\eta,z)$ of  \cite{Abram} for $\eta=0$.
With the help of standard expansions for the Bessel functions 
\cite{Abram}
one can obtain the following representations for the functions $F_\lambda(z)$ and $G_\lambda(z)$, which are needed below to analyze the low-energy behavior of solutions,
\begin{align}
F_\lambda(z)= z^{\lambda+1}\sum_{k=0}^{\infty} f_k z^{2k}  \equiv z^{\lambda+1} 
{\tilde  F}_\lambda(z^2), \label{F0}
\\ 
G_\lambda(z)= z^{-\lambda}\sum_{k=0}^{\infty} g_k z^{2k}  \equiv z^{-\lambda} 
{\tilde  G}_{\lambda}(z^2).  \label{G0}
\end{align} 
The coefficients  $f_k$ and $ g_k$ for each $\lambda$ do not depend on  $z$. 
The following asymptotic representations also hold for  $|z| \to \infty$
\begin{align}
F_\lambda(z) \sim \sin(z-\lambda\pi/2) \nonumber 
, \\
G_\lambda(z) \sim \frac{1}{\cos \lambda \pi}\cos(z+\lambda\pi/2).
\label{G-infty}
\end{align}
The right hand side of  (\ref{G-infty}) takes the standard form  $\cos(z-\ell\pi/2)$ at integer value of  $\lambda=\ell \ge 0$ .   

The parameters $a_1(p)$ and $b_1(p)$ are determined from the continuity conditions on 
the solution $\phi(r,p)$ and its derivative at the point $r=R$ and are given by the formulas
\begin{align}
a_1(p)=-\frac{W(F_\lambda,G_\lambda)}{W(G_\lambda,F_\ell)}, \ b_1(p)=-\frac{W(F_\lambda,F_ \ell)}{W(G_\lambda,F_\ell)}.
\label{a1b1}
\end{align}
In these formulas, the Wronskians of the form $W(f,g)=f {\dot g}- {\dot f}g$
are calculated at $r=R$. Here and below, the dot denotes the derivative with respect to the variable $r$,
for example, ${\dot f}= \frac{d}{dr}f$.	

Let us construct the second linearly independent with $\phi(r,p)$ solution $\gamma(r,p)$ of the equation (\ref{SE-V^R}), irregular at the point $r=0$. Let us put 
\begin{align}
&\gamma(r,p)= a_2(p)F_\ell(qr)+b_2(p) G_\ell(qr), \ \ \ 0\le r\le R, \nonumber 
\\
&\gamma(r,p)= G_\lambda(pr), \ \ \ r\ge R.  \nonumber 
\end{align}
As before, matching solutions at the point $r=R$ leads to the following values for $a_2$ and $b_2$
\begin{align}
a_2(p)=-\frac{W(G_\lambda,G_\ell)}{W(G_\ell,F_\ell)},\, b_2(p)= \frac{W(G_\lambda,F_ \ell)}{W(G_\ell,F_\ell)},
\label{a2b2}
\end{align}
where the Wronskians are calculated at the point $r=R$.

In a standard way, the solutions $\phi(r,p)$ and $\gamma(r,p)$ allow us to construct the Green's function $\Gamma(r,r',p)$, which gives a solution to the inhomogeneous equation
\begin{equation*}
\left[-\frac{d^2}{dr^2} +\frac{\ell(\ell+1)}{r^2}+V^R(r)-p^2 \right] \Gamma(r,r',p)=\delta(r-r'),  
\end{equation*} 
 in the form 
 \begin{equation*}
 \Gamma(r,r',p)=-\frac{\phi(r_{<},p)\gamma(r_{>},p)}{W(\phi,\gamma)}.  
\end{equation*}
Here  $r_>(r_<)=\max(\min)[r,r']$. 
The Wronskian in the denominator does not  depend on 
 $r$ and is given by the formula  
$$
W(\phi,\gamma)=-a_1(p)b_2(p) q.
$$ 
Using the Green's function $\Gamma(r,r',p)$, we proceed to the second stage of solving the problem and transform (\ref{SE-R}) to the integral L-S   equation 
\begin{equation}
\psi(r,p)=\phi(r,p)-\int_0^R dr' \, \Gamma(r,r',p)V_R(r')\psi(r',p). 
\label{LS}
\end{equation}	
The structure of the equation (\ref{LS}) is such that to find $\psi(r,p)$ on the entire interval 
$0\le r< \infty$, it suffices to find the solution of (\ref{LS}) for $0\le r \le R$.
Indeed, restricting (\ref{LS}) to the interval $0\le r \le R$, we obtain a closed integral equation for the function $\psi_R(r,p)=\psi(r,p)$ for $0\le r \le R$
\begin{equation}
{\psi}_R(r,p)=a_1(p) F_{\ell}(qr)-\int_0^R dr' \, \Gamma(r,r',p)V_R(r'){ \psi}_R(r',p). 
\label{LS_r}
\end{equation}
At the same time, the equation (\ref{LS}) with $R\le r<\infty$ now becomes just a formula for calculating $\psi(r,p)$ from ${\psi}_R(r,p) $
\begin{align}
\psi(r,p)= F_\lambda(pr)+G_\lambda(pr) {\cal K}_{\lambda\ell}(p), 
\label{psi-R-infty}
\end{align} 
where  ${\cal K}_{\lambda\ell}(p)$ is given by the integral  
\begin{align}
{\cal K}_{\lambda\ell}(p)= b_1(p)-\frac{1}{b_2(p)q}\int_0^R dr'\, F_\ell(qr')V_R(r'){\psi}_R(r',p). 
\label{K-hat}
\end{align} 
Thus, to solve the scattering problem on the entire interval $0\le r<\infty$, it suffices to solve the equation for $\psi_R(r,p)$ on the interval $0\le r\le R$.
In section \ref{SLS} it is shown that the integral equation (\ref{LS_r}) can be reduced
to the Volterra equation, the solution of which  is given by an iterative expansion \cite{DeAlfaro}.
The last circumstance guarantees the existence and uniqueness of a solution to the equation for 
${ \psi_R(r,p)}$, which uniquely determines the required solution of the scattering problem $\psi(r,p)$ on the entire interval $0\le r<\infty$.

The resulting representation (\ref{psi-R-infty}) can now replace the asymptotic condition (\ref{psi-as}) in the formulation of the scattering problem. Note in this connection that the representation (\ref{psi-R-infty})
is valid starting from $r=R$, and at the same time, the quantity $pR$ can be not large, moreover it can take arbitrary small value for $p\to 0$.  
For the domain of the physical asymptote of the wave function (\ref{psi-as}), $pr\to\infty$ is required for $r\ge R$. In this asymptotic region, $\psi(r,p)$ takes the form
\begin{align*}
\psi(r,p)\sim \sin(pr-\lambda\pi/2)+  
\frac{\cos(pr+\lambda\pi/2)}{\cos \lambda \pi}{\cal K}_{\lambda\ell}(p).
\end{align*}
In turn, the right hand side of the last relation is transformed into the form of the standard physical asymptotics (\ref{psi-as}) of the solution of the scattering problem
\begin{align*}
\psi(r,p)\sim &\left[ \sin(pr-\ell\pi/2)+\cos(pr-\ell\pi/2) K_{\ell}(p)\right] A_{\lambda\ell}, 
\end{align*}
where the physical  $K$-matrix is given by 
\begin{align}
K_{\ell}(p)= 
\frac{\cos(\lambda\pi)\sin((\ell-\lambda)\pi/2)+{\cal K}_{\lambda\ell}(p)\cos((\ell+\lambda)\pi/2)}
{\cos(\lambda\pi)\cos((\ell-\lambda)\pi/2)-{\cal K}_{\lambda\ell}(p) {\sin((\ell+\lambda)\pi/2)}}, 
\label{K}
\end{align}
and the amplitude $A_{\lambda\ell}$ is given by the formula 
$$
A_{\lambda\ell}= \cos((\ell-\lambda)\pi/2)-
{\cal K}_{\lambda\ell}(p)\frac{\sin((\ell+\lambda)\pi/2)}{\cos(\lambda\pi)}. 
$$
So, we have constructed a solution to the scattering problem (\ref{SE})-(\ref{psi-as}) and obtained a representation (\ref{K})  for the $K$-matrix. The next section analyzes the low-energy
behavior of $K_{\ell}(p)$ as $p\to 0$.

\section{Low-energy behavior of  $K$-matrix}
The dependence of the $K$-matrix (\ref{K}) on $p$ is completely determined by the function
${\cal K}_{\lambda \ell}(p)$ from (\ref{K-hat}). 
The latter depends on $p$ through the combination $pR$, on which the functions $F_\lambda$ and $G_\lambda$ depend, and also through the combination $qR$, on which depend the functions 
$F_\ell$ and $G_\ell$. 
Since $q^2=p^2 + V_0$, then when $p\to 0$ the variable $q$ becomes independent of $p$. For this reason, the functions $F_\ell$ and $G_\ell$ in the leading  order can be considered as independent of $p$. Thus, all dependence on $p$ will be concentrated in the functions $F_\lambda$ and $G_\lambda$.
Therefore, in this section we introduce the variable $z=pR$ and study the behavior of the $K$-matrix as a function of $z$ as $z\to 0$.
Representations for $b_1$ and $b_2$ needed in (\ref{K-hat}) as $z\to 0$
are obtained from expressions found for them in (\ref{a1b1}) and (\ref{a2b2}) and from definitions (\ref{F0}) and (\ref{G0})
\begin{align}
b_1= &z^{2\lambda+1} {\tilde  b}_1=
-z^{2\lambda+1}\frac{f_0{\dot F}_\ell - \frac{(\lambda+1)f_0}{R}F_\ell}
{g_0{\dot F}_\ell + \frac{\lambda g_0}{R}F_\ell}(1+O(z^2)), 
\label{b11} 
\end{align}
\begin{align}
b_2=z^{-\lambda}{\tilde  b }_2=
z^{-\lambda}
\frac{g_0{\dot F}_\ell+ \frac{\lambda g_0}{R}F_\ell}
{q 
}
(1+O(z^2)).
\label{a22}
\end{align}
In these formulas, as before, the dot denotes derivatives with respect to the variable $r$, and the functions $F_\ell$ and ${\dot F}_\ell$ are calculated at the point $r=R$.
To estimate the integral term in (\ref{K-hat}), let us consider in more detail the integral equation (\ref{LS_r}) for the function $\psi_R$ on the interval $0\le r\le R$
\begin{align}
{\psi}_R(r,p)=&
a_1F_\ell(qr)
-\frac{F_\ell(qr)}{q}\int_r^R  \left[\frac{a_2}{b_2}F_\ell(qr') +G_\ell(qr')  \right] 
V_R(r'){\psi}_R(r',p) dr' - \nonumber \\
-&\left[\frac{a_2}{b_2}F_\ell(qr)+G_\ell(qr)\right] \frac{1}{q}\int_0^r F_\ell(qr') V_R(r') {\psi}_R(r',p)\,dr'. 
\label{psi-rr}
\end{align}
For $a_1$ and $a_2/b_2$, similarly to what was done above,  we obtain the following representations
\begin{align}
a_1=\frac{z^{\lambda+1}}{R}{\tilde  a}_1= 
\frac{z^{\lambda+1}}{R}\frac{(2\lambda +1)f_0}{{\dot F}_\ell +\frac{\lambda}{R}F_\ell}(1+O(z^2)), \label{a1}    \\
\frac{a_2}{b_2}=-\frac{{\dot G}_\ell+\frac{\lambda }{R}G_\ell}{{\dot F}_\ell+\frac{\lambda}{R}F_\ell}(1+O(z^2)) \label{b/a}, 
\end{align}
 where functions  $F_\ell$, $G_\ell$  and their derivatives are calculated at the point $r=R$.  
 It can be seen from the formulas (\ref{a1}) and (\ref{b/a}) that $a_2/b_2$ becomes a constant as $z\to 0$, and also the  renormalized value of ${\tilde a}_1 = Rz^{-\lambda-1}a_1$ becomes  a constant. 
 Hence,  the renormalized function
 ${\tilde \psi}_R(r,p)=Rz^{-\lambda-1}{\psi_R(r,p)}$ will obey  the renormalized equation (\ref{psi-rr}) in which $ a_1$ is replaced by ${\tilde a}_1$. 
Since in the equation thus obtained the driving  term and the kernel become quantities  independent of $z$ as $z\to 0$, then its solution ${\tilde \psi}_R(r,p)$ will have the same property. As a result, we arrive at the following low-energy behavior of the original solution
${ \psi}_R(r,p)$ as  $z\to 0$ 
\begin{equation}
{\psi}_R(r,p) =  \frac{z^{\lambda+1}}{R} {\tilde \psi}_R(r,p),
\label{psi-zR}
\end{equation}
where  ${\tilde \psi}_R(r,p)$ becomes independent of  $z$ as $z\to 0$. 
The formulas (\ref{b11}), (\ref{a22}), and (\ref{psi-zR}) provide the following final factorized representation for ${\cal K}_{\lambda\ell}(p)$, in which we return to the original variable $p$,
\begin{align}
{\cal  K}_{\lambda\ell}(p) = (pR)^{2\lambda+1} {\tilde {\cal K}}_{\lambda\ell}(p), \label{K-hh}\\
{\tilde {\cal K}}_{\lambda\ell}(p)= {\tilde  b}_1-\frac{1}{q{\tilde b}_2R} \int_0^R dr' F_\ell(qr')V_R(r'){\tilde \psi}_R(r',p). \label{K-t-final}
\end{align}
Here the renormalized ${\tilde {\cal K}}_{\lambda\ell}(p)$ becomes $p$-independent as $pR\to 0$. The resulting formula (\ref{K-hh}) after substitution into (\ref{K}) gives a generalization  of the representa\-tion for the threshold behavior of the $K$-matrix to the case of dipole interaction. Note that for $\alpha^2=0$, the formulas obtained above lead to the standard threshold behavior for the $K$-matrix. Indeed, in this case $\lambda=\ell$, which implies the equality $K_\ell(p)={\cal K}_{\ell\ell}(p)$ and, accordingly, the formula (\ref{K-hh}) reproduces the threshold behavior for 
short-range potentials
$K_\ell(p)\propto  (pR)^{2\ell+1}$.

Formulas (\ref{psi-zR}) and
(\ref{K-hh}), (\ref{K-t-final}) are the main results of our study of low-energy 
scattering off the potential with dipole interaction at long distances.
In contrast to the analogous result  of \cite{GD-1, GD-2}, the formulas (\ref{K-hh}), (\ref{K-t-final}) give explicit representations for the $K$-matrix in terms of the wave function. The latter, in turn, is given by the solution of the Lippmann-Schwinger integral equation (\ref{LS}), which seems to be an important advantage of the formalism developed in this paper.

\section{Analysis of the low-energy scattering in the system of electron, positron and  antiproton} 

\begin{figure}
\includegraphics[width=0.5\textwidth]{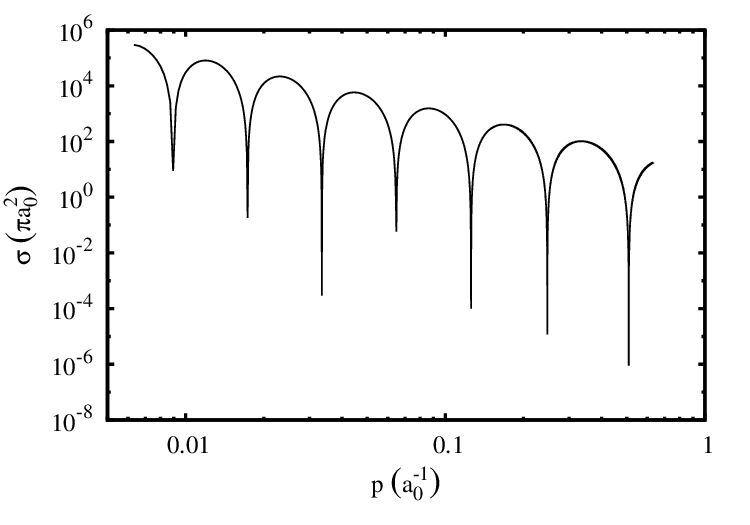}
\caption{Cross section  (in units  of $\pi a_0^2$) of scattering off the potential $V^R$ with parameters  (\ref{vr-param}). 
}
\label{CS-fig}
\end{figure}

In this section, we use the obtained results  to analyze the scattering of an antiproton by positronium $\text{Ps}$ (a bound state of an electron and a positron).
As mentioned in the introduction, it is this system that attracts attention in connection with experiments on antimatter.
The polarization potential between an antiproton and an electron-positron pair arises when $\text{Ps}$ is in an excited state. Let us consider the case of the first excited state with the principal quantum number $n=2$, in which the orbital momentum of the pair can take the values $\ell=0,1$.
After projecting the potentials between the antiproton and the $\text{ Ps}$ constituents  onto the wave functions of the target ($\text{Ps},\, n=2$), a matrix effective polarization potential arises, which, at a sufficiently large distance $r$ between $\text{ Ps}$ and $\bar p$ has the form 
\begin{equation*}\label{pol-pot}
V_{\text{Ps}-\bar{p}}(r)=\frac{1}{r^2}
\left( 
\begin{array}{cc}
0& 23.9739 
\\  
23.9739& 0 
\end{array} 
\right). 
\end{equation*}  
Here and in what follows, atomic units are used for energy and distance with the atomic unit of length $a_0$.
Together with the centrifugal interaction $V_{\text{Ps}-\bar{p}}$ forms the resulting matrix potential
\begin{equation*}
V_{\text{eff}}(r) = 
\frac{1}{r^2} 
\left( 
\begin{array}{cc}
0& 23.9739 
\\  
23.9739& 2
\end{array} 
\right). 
\end{equation*}
The solution of the two-channel Schrödinger equation with such a potential is carried out using the diagonalization of $V_{\text{eff}}$ 
\begin{equation*}
U^{\dagger}V_{\text{eff}}(r)U = 
\frac{1}{r^2} 
\left( 
\begin{array}{cc}
24.9947& 0 
\\  
0& - 22.9947 
\end{array} 
\right) 
\label{V-eff-d}
\end{equation*}
and followed by the parametrization  \cite{GD-1, GD-2, Burke} of the form 
\begin{equation*}
U^{\dagger}V_{\text{eff}}(r)U = \frac{1}{r^2}\Lambda(\Lambda+1) .
\label{V-eff-d-lambda}
\end{equation*}
Here, the diagonal matrix $\Lambda$ is given by 
\begin{equation*}
\Lambda=
\left( 
\begin{array}{cc}
\lambda_1& 0 
\\  
0& \lambda_2
\end{array} 
\right) = 
\left( 
\begin{array}{cc}
4.52441& 0 
\\  
0& -0.5 +\text{i} 4.76914 
\end{array} 
\right),  
\label{Lambda}
\end{equation*}
where $\text{i}=\sqrt{-1}$. 
The complex value of $\lambda_2$ corresponds to the situation when a sufficiently intensive polarization attractive potential arises in a given channel at large distances.
It is in this situation that the oscillations predicted in \cite{GD-1, GD-2} appear in low-energy scattering. Indeed, according to
(\ref{K-hh}, \ref{K-t-final}) the contributions to the $K$-matrix from the asymptotic part of the interaction and from the interaction at medium and small distances $V(r)$ have the same dependence on the relative momentum $p$
\begin{equation*}
(pR)^{2\lambda +1}. 
\label{pR}
\end{equation*}
In the case $\lambda=\lambda_2$ this dependence produces  an oscillating factor 
\begin{equation*}
\exp\{\text{i} 2(\Im{\text m}\lambda_2)\ln(pR)  \}. 
\label{exp}
\end{equation*}
To illustrate the arising oscillations, consider the solution of a model scattering problem with the potential $V^R$ with parameters corresponding to the real parameters of the effective
$\text{Ps}-\bar{p}$ potential
\begin{equation*}
\label{vr-param}
\ell = 1,\ \ell(\ell+1)- \alpha^2= -22.9947,\ R = 6.0\, a_0.
\end{equation*} 
Fig.~\ref{CS-fig} shows a logarithmic in the $p$ variable plot
 of the scattering cross section calculated on the basis of the representation for the $K$-matrix (\ref{K}) by the formula
$$
\sigma_1 = \frac{3\pi}{p^2}\left|\frac{2 \text{i} K_1(p)}{1-\text{i} K_1(p)}\right|^2. 
$$ 
The logarithmic nature of the position of the maxima (minima) of the cross section, dictated by the dependence (\ref{exp}), is clearly visible.

Of particular interest is the behavior of the wave function which is  the solution of the scattering problem (\ref{SE})-(\ref{psi-as}). Squared wave functions normalized by the asymptotic condition (\ref{psi-as}) with $A_\ell=1$ for two momenta $p=0.006\, a_0^{-1}$ and $ p=0.02\, a_0 ^{-1}$ have the form shown in Fig.~2.
\begin{figure}
\includegraphics[width=0.5\textwidth]{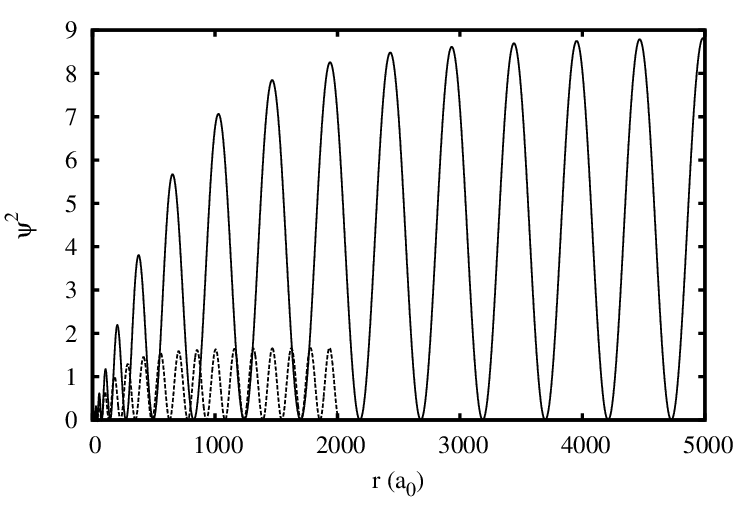}
\caption{Squared wave functions normalized by the asymptotic condition~(\ref{psi-as}) with $A_\ell=1$ for the potential $V^R$ with parameters~(\ref{vr-param}) for different values of $p $: solid line corresponds to $p=0.006\, a_0^{-1}$, dashed line corresponds to $p = 0.02 \, a^{-1}_0$.
}
\label{WF-fig}
\end{figure}
The following values of the $K$-matrices correspond to taken momenta: 
${\cal K}_{\lambda\ell}(0.02)=3.12\cdot10^5+\text{i} 1.58\cdot10^6$, $K_ \ell(0.02)=-0.822$,
${\cal K}_{\lambda\ell}(0.006)=-1.25\cdot10^6+\text{i}1.01\cdot10^6$, $K_\ell(0.006)=-2.82$.
It can be seen that  amplitudes of oscillation become constant at very large distances, $r> 2000\, a_0$ for $p=0.02\, a_0^{-1}$ and $r>5000\, a_0$ for $p=0.006\, a_0^{-1}$, respectively.
These distances specify  the regions in which the wave functions approach  the physical asymptotics (\ref{psi-as}). At the same time, the wave functions become equal to the superposition (\ref{psi-R-infty})
of functions $F_\lambda$ and $G_\lambda$ for $r\ge 6 \, a_0$. Thus, choosing the representation (\ref{psi-R-infty}) as the asymptotic boundary condition instead of (\ref{psi-as}) reduces the size of the domain in which the Schr\"odinger equation needs to be solved by three orders of magnitude on average.

\section{Analysis of integral equation  (\ref{LS_r}) }\label{SLS}
Here we reduce the Fredholm-type integral equation (\ref{LS_r}) to a Volterra-type equation, to which the results of \cite{DeAlfaro} can be directly  applied.
Using the explicit form of the Green's function $\Gamma(r,r',p)$, we write the equation (\ref{LS_r}) as
\begin{align}
\psi_R(r,p)=&\phi(r,p)+ \frac{1}{W(\phi,\gamma)}\int_0^r dr' \phi(r',p)\gamma(r,p)V_R(r')\psi_R(r',p)+  \nonumber \\
+&\frac{1}{W(\phi,\gamma)}\int_r^R dr' \phi(r,p)\gamma(r',p)V_R(r')\psi_R(r',p). 
\label{eq1}
\end{align}
We transform the integral in the last term of this equation by adding and subtracting the integral over the interval $[0,r]$ with the same integrand. 
As a result, after obvious transformations, we obtain the equation
\begin{align*}
\psi_R(r,p)=\phi(r,p)\left[1+\frac{1}{W(\phi,\gamma)}\int_0^R dr' \gamma(r',p)V_R(r')\psi_R(r',p) \right] + \nonumber \\
+\frac{1}{W(\phi,\gamma)}\int_0^r dr' 
\left[\phi(r',p)\gamma(r,p)-\phi(r,p)\gamma(r',p)\right]V_R(r')\psi_R(r',p). 
\end{align*}
Introducing a new unknown function 
$$
{\hat \psi}_R(r,p)= \psi_R(r,p)
\left[1+\frac{1}{W(\phi,\gamma)}\int_0^R dr' \gamma(r',p)V_R(r')\psi_R(r',p) \right]^{-1}, 
$$
we obtain for it the Volterra integral equation on the interval $[0,R]$
\begin{align}
{\hat \psi}_R(r,p)&=\phi(r,p) 
+ \nonumber \\
&+\frac{1}{W(\phi,\gamma)}\int_0^r dr' 
\left[\phi(r',p)\gamma(r,p)-\phi(r,p)\gamma(r',p)\right]V_R(r'){\hat \psi}_R(r',p). 
\label{3}
\end{align}
The equalities $\phi(r,p)=a_1F_\ell(qr)$ and $\gamma(r,p)=a_2F_\ell(qr)+b_2G_\ell(qr)$ hold on the interval $[0,R]$. As a result, the driving term in (\ref{3}) differs only by a constant from the driving  term of the integral equation (3.29) from \cite{DeAlfaro}. The same applies to the kernel in (\ref{3}), which differs from the kernel  
of the same equation from \cite{DeAlfaro} only by an insignificant term proportional to $F_\ell(qr)F_\ell(qr')$. Under these conditions, we have the right to use the result of \cite{DeAlfaro} as applied to (\ref{3}) to prove the existence and uniqueness of a solution of this equation, which is given by an iterative series. The solution of  the original equation (\ref{eq1}) is now expressed by the formula
$$
{\psi}_R(r,p)= {\hat \psi}_R(r,p)
\left[1-\frac{1}{W(\phi,\gamma)}\int_0^R dr' \gamma(r',p)V_R(r'){\hat \psi}_R(r',p) \right]^{-1}. 
$$

\section{Conclusion}
In conclusion of this work, we formulate the main results. On the basis of the potential splitting method, a formalism is developed  that explicitly takes into account the induced dipole interaction.
To solve the scattering problem using the potential splitting method, an integral equation of the Lippmann-Schwinger type is constructed.
The structure of this equation is such that its solution for the inner range  makes it possible to construct a solution in the outer range 
using a simple quadrature. This property favorably distinguishes the formalism developed in this paper from the $R$-matrix method, in which, in order to find a solution in the inner region, it is required to match it with the asymptotic solution at the interface between the inner and outer regions. Moreover, the proportionality coefficient is the unknown $R$-matrix, which must also be determined in the process of solution.

The Lippmann-Schwinger integral equation for the wave function
made it possible to obtain an explicit representation for its low-energy behavior.
By  solving this integral equation, an explicit integral representation is obtained for the $K$-matrix of the scattering problem in terms of the wave function.
Based on this integral representation, the low-energy behavior of the $K$-matrix is rigorously investigated, confirming the result of \cite{GD-1, GD-2}.

To illustrate the extremely slow approach of the solutions of the scattering problem with dipole interaction to its asymptote  we analyze 
the low-energy behavior of the wave function and scattering cross section for the $\text{Ps}-\bar{p}$ scattering channel, in which the characteristic oscillations of the scattering cross section predicted in \cite{GD-1, GD-2} arise.  The results obtained will undoubtedly be useful for  solving realistic  three-particle problems with charged particles, in which explicit treatment of  the dipole interaction will make it possible to overcome the main computational difficulty associated with the extremely slow approach  of three-particle wave functions and their Faddeev components to the asymptotes 
\cite{Hu} - \cite{Grad3}.

\bigskip
\bigskip
\noindent
{\bf \large{Acknowledgment} }\\ 
This work was carried out within the framework of the Russian Science Foundation project No. 23-22-00109 using the equipment of the Resource Center "Computer Center of St. Petersburg State University" (http://cc.spbu.ru).

\end{document}